# Rheology of sedimenting particle pastes

Abdoulaye Fall, Henri de Cagny, Daniel Bonn

Van der Waals-Zeeman Institute, IoP, University van Amsterdam, Science Park 904, 1098XH Amsterdam, The Netherlands

Guillaume Ovarlez

Université Paris Est-Institut Navier, 2 allée Kepler, 77420 Champs-sur-Marne, France

Elie Wandersman[1], Joshua A. Dijksman[2], Martin van Hecke

Kamerlingh Onnes Laboratory, LION, Universiteit Leiden, P.O. Box 9504, 2300 RA Leiden, The Netherlands

1. Also at: Laboratoire Jean Perrin, FRE 3231, CNRS / Université Pierre et Marie Curie, Ecole Normale Supérieure, 24 rue Lhomond, 75005 Paris, France
2. Also at: Physics Department, Duke University, Science Dr, Box 90305 Durham NC 27708 USA.

Abstract

We study the local and global rheology of non-Brownian suspensions in a solvent that is not density-matched, leading to either creaming or sedimentation of the particles. Both local and global measurements show that the incomplete density matching leads to the appearance of a critical shear rate above which the suspension is homogenized by the flow, and below which sedimentation or creaming happens. We show that the value of the critical shear rate and its dependence on the experimental parameters are governed by a competition between the viscous and gravitational forces, and present a simple scaling model that agrees with the experimental results from different types of experiments (local and global) in different setups and systems.

# 1. Introduction

Dense suspensions exhibit very rich behavior: some suspensions exhibit a yield stress [Husband et al. 1993, Ancey et al. 1999, Prasad et al. 1995, Huang et al. 2005, 2007 Ovarlez et al. 2006], many show shear banding [Barentin et al. 2004, Huang et al. 2005, Ovarlez et al. 2006, Fall et al. 2009, Schall and van Hecke 2010], and some show shear thickening [Fall et al. 2008, 2012, Brown and Jaeger 2009, 2011, 2012, Brown et al. 2010, Cheng et al. 2011], normal stresses [Morris et al. 1999, Zarraga et al. 2000] and shear-induced migration [Morris et al. 1999, Ovarlez et al. 2006, Fall et al. 2010] that remain incompletely understood. A crucial role is played by the density mismatch between the particles and fluid [Fall et al. 2009]. Non-density matched suspensions, such as sand in water, share features with both dry granular media and density matched suspensions [Dijksman et al. 2010].

On the one hand, dry granular media are collections of non-Brownian particles that interact only when in contact. Their interaction is dissipative, dominated by inelastic collisions and frictional contacts. In spite of these seemingly simple interactions, their collective behavior is very rich and complex, and has attracted much attention in recent years [Jaeger et al. 1996, GDR MiDi 2004]. With the exception of very rapid flows [Goldhirsch 2003], in most cases friction dominates. This then suggest that the resistance to flow, a subject of considerable fundamental interest and practical importance, is also frictional. While there remain, in particular for very slows, several open questions [Fenistein et al. 2003, 2004, 2006], the relation between stresses and strain rate have been studied in much detail, and there is little doubt that they indeed are very similar to dry friction: the shear stress is proportional to the normal stress, only weakly varies with the strain rate, and reaches a finite yield stress in the limit of vanishing strain rate [Jop et al. 2005, GDR MiDi 2004, Dijksman et al. 2010, 2011].

On the other hand, density matched suspensions behave as Newtonian liquids, with an effective viscosity which grows rapidly with the packing fraction [Bonnoit et al. 2010]. It has recently been suggested [Boyer et al. 2011] that it is possible to reinterpret this growth of the viscosity, observed at constant packing fraction, as arising via the growth of the normal stresses [Boyer et al. 2011]. Nevertheless, the rheology is very different from dry granular media – the stresses are proportional to the strain rate and thus vanish when the strain rate goes to zero. To observe this absence of a yield stress, it is crucial that the density matching is very precise – any density mismatch will lead to a finite yield stress [Fall et al. 2009]

For the general case of non-density matched suspension, recent experiments point to the existence of (at least) two distinctly different flow regimes. There is a slow flow regime in which the contacts between particles are essentially frictional and the rheology is similar to that of dry granular media, while for fast enough flows, the particles are resuspended, lose contact, and the rheology becomes Newtonian [Fall et al., 2009, Moller et al., 2009 Dijksman et al. 2010]. The crucial question then becomes: what is the critical strain rate where the crossover between these two regimes takes place?

Here we probe the flow of a range of non-density matched suspensions by a combination of MRI, stress controlled rheometry and rate controlled rheometry. In all cases, we observe a critical strain rate: stress controlled rheometry evidences this critical strain rate via a viscosity bifurcation, while rate controlled rheometry (and MRI) evidence the critical strain in the rheological curves. We will show that the crossover strain rate is proportional to the density mismatch and granular pressure, thus evidencing that it can be understood to arise from a balance of static, frictional stresses to the dynamic, viscous stresses, and introduce a scaling model which is in full agreement with our data.

## 2. Rheological curves and critical flow rates

In a first series of experiments, we probe the local and global rheology of dense suspensions composed of non-colloidal monodisperse spherical particles (polystyrene beads, diameter $40\mu m$, polydispersity < 5%, density 1050 Kg.m$^{-3}$) immersed in a Newtonian fluid (water + NaI) at a volume fraction $\varphi$ of 60%. By varying the salt concentration, we can perfectly match the solvent and particle densities but also reach a wide range of density differences between the solvent and particles. In all cases where we have a density difference, we keep a relatively high salt concentration, implying that the particles cream rather than sediment [Fall et al. 2009].

MRI rheometry is performed in a wide-gap Couette geometry ($R_i$ = 41.5 mm, $R_e$ = *60* mm, H = 110 mm). Local velocity and concentration profiles in the flowing sample were obtained through magnetic resonance imaging (MRI) techniques described in detail in [Rodts et al. 2004, Bonn et al. 2008]. For all experiments, in order to avoid slip at the walls, sandpaper of roughness equivalent to that of the particles is glued on the walls; we checked on the velocity profiles that there is no observable slip. We investigated the stationary flows for inner cylinder rotational velocity $\Omega$ ranging between $10^{-3}$ and 0.16 rps.

By combining the local shear stress and local strain rate, we will determine the local rheology of our suspensions. The stress is found by measuring the torque necessary to turn the inner cylinder and using the well-known $1/r^2$ variation of the shear stress in a Couette geometry (which follows from momentum balance). The shear rate can be deduced from the velocity $v(r)$ profile as $\dot{\gamma} = -r\partial(v/r)/\partial r$ - we therefore compute for each distance $r_i$ the shear rate from the values of the velocity between two points:

$$\dot{\gamma}(r_i) = -r_i(v_{i+1}/r_{i+1} - v_{i-1}/r_{i-1})/(r_{i+1} - r_{i-1}) \qquad (1)$$

In Fig. 1, we show the resulting local rheological curves for both a density matched ($\Delta\rho = 0$) and non-density matched ($\Delta\rho = 150\ Kg/m^3$) suspension. The density-matched material flows as a Newtonian fluid (a slope of 1 on the log-log scale implies a constant viscosity) whereas a yield stress of about 3.2 *Pa* is observed for the non-density matched system. As the figure shows, the local rheology data for the non-matched system can be fitted by a Bingham model: $\sigma = (3.2 + 5.1\dot{\gamma})Pa$ although the range of stresses is small.

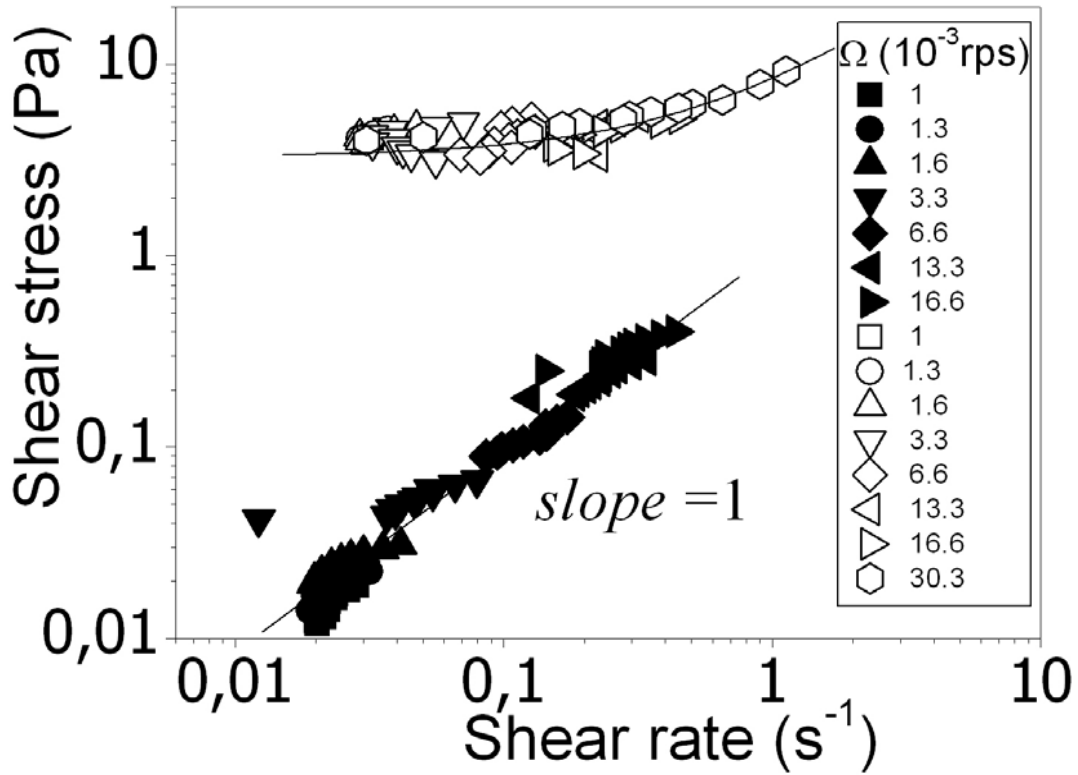

Fig. 1: Constitutive law from the MRI measurements in the density-matched (closed symbols) suspension with shows a Newtonian behavior and in the mismatched suspension (open symbols) where a yield stress behavior is observed. The continuous line is a Bingham model. The line through the Newtonian behavior gives a slope of unity. Up to the highest strain rates shown here, no particle migration is observed and thus the volume fractions are the same in both cases, i.e., 60%. The system studied here are 40 μm Polystyrene beads suspended in a mixture of water and NaI; the latter allows to tune the density difference.

In a second set of experiments, we probe the macroscopic rheology of a non-density matched suspension ($\Delta\rho = 150\ Kg/m^3$) with a vane-in-cup geometry (inner cylinder radius $R_i$ = 12.5 *mm*, outer cylinder radius $R_e$ = 18.5 *mm*, height H = 45 *mm*) on a commercial rheometer (Bohlin C-VOR 200). Since the creaming is time-dependent, so is the rheology of such suspensions. Applying a sufficiently high shear stress leads to a homogenization of the system, whereas for small shear stresses, creaming will generate a yield stress - the viscosity will then increase in time without bound, in turn implying an infinite viscosity in steady state. Both the viscosity and the yield stress are thus time dependent, and this leads to a *viscosity bifurcation* in the system as a function of the applied stress [Moller et al. 2006, 2008, Fall et al. 2009] – see Fig. 2a

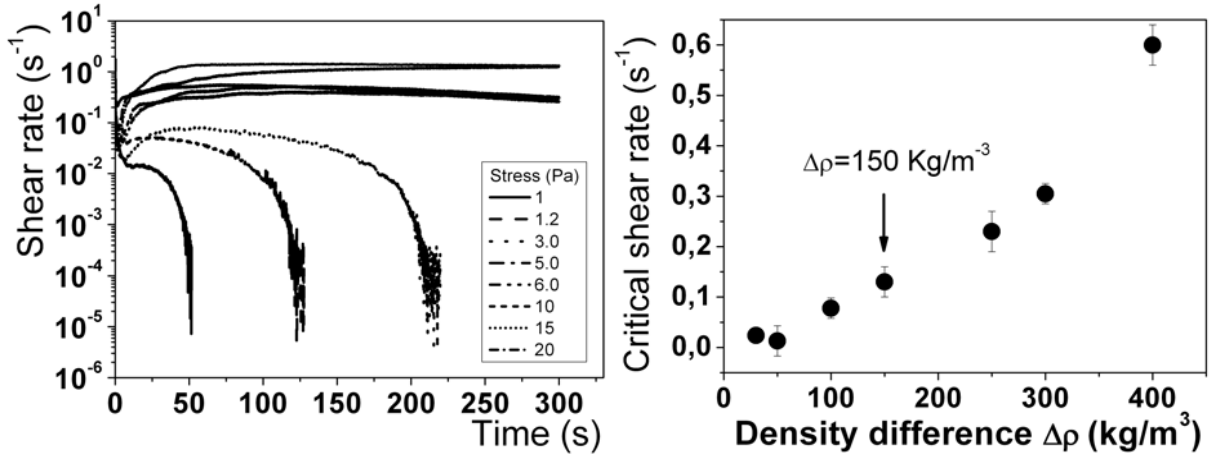

Fig. 2: (a) Shear rate *vs.* time for different applied shear stresses in a mismatched suspension ($\Delta\rho$=150 Kg/m$^3$) (b) Critical shear rate as a function of the density difference [Fall et al. 2009].

As was shown in [Fall et al. 2009], the viscosity bifurcation allows one to define a critical shear rate, as the lowest shear rate for which a steady state flow can be sustained; this critical rate is found to vary linearly with density difference (Fig 2b). This suggests that this critical strain rate is governed by a simple stress balance, where the viscous stresses $\approx \eta\dot{\gamma}$ are balanced by a compressive stress due to gravity. From dimensional analysis it follows that $\eta\dot{\gamma} \propto \Delta\rho g L$, with **L** a length scale. What is this length scale **L**? Fall et al. (2009) suggested to take the particle radius **R** for this, leading to $\eta\dot{\gamma}_c = \Delta\rho g R$. This was observed to give a reasonable estimate of the critical shear rate of Fig. 2(b). However it is unclear why the particle radius and not the suspension height should enter this stress balance.

To probe the rheology of non-density matched suspensions in more detail, and to probe the pressure scale that sets the critical flow rate, we performed a third series of experiments in a Couette geometry, this time varying the filling height of different non density matched granular suspensions. The granular suspension consists of 4.6 mm PMMA beads immersed in Triton X-100 (viscosity η = 0.23 Pa.s, $\Delta\rho = 107 \, kg/m^3$ at 25 °C), a similar granular suspension as the one studied in [Dijksman et al. 2010]. The rheological measurements are performed in a 'home designed' Couette geometry: the suspension is stored in an open-topped cubic box (width 150 mm). A toothed metallic cylinder (radius $R_s$ = 25 mm, tooth dimension = 2.5 mm) connected to a rheometer (*Anton Paar MCR501*) is used to shear the suspension at a constant rotation rate $\Omega$ (from $\Omega=10^{-4}$ to 1 rps). The filling height of the material is determined by weighing the particles ($H/R_s$ is varied from 0.5 to 2 with an absolute error of $d/R_s=0.2$). The Couette cell is then filled by adding a known weight of particles. The system is manually stirred, the top is flattened, and the filling height is re-measured with a ruler. The estimated initial volume fraction is $\phi \approx 0.6$. For all the experiments, there is at least a 1 cm fluid layer over the top of the granular suspension. Experiments consisted of a logarithmic ramping of the rotation speed ($3.10^{-4}$ to 0.3 rps in 30 steps) along with the time interval each speed is held. This time ($10^3$ s to 1 s) is chosen so that each rotation rate is held for 0.3 rotations. We do not observe significant effects in the results if the duration of the time intervals is made longer. In addition, experiments where the velocity is reversed (from 0.3 to $3.10^{-4}$) present small hysteresis only. The system is thus in a steady-state at each rotation rate, and has the time to dilate at each step.

Rheological measurements are presented on Fig 3a, for various filling heights. We used a Herschel-Bulkley type model to fit the data

$$T = T_0 + K\Omega^b \quad (2)$$

As reported earlier in [Dijksman et al. 2010] with a similar suspension sheared in the Split-Bottom geometry, if we fit Eq. (2) to the data we find b ≈ 1 for all $H/R_s$, i.e., a Bingham type of rheological law is measured, consistent with the data presented in Fig. 1.

A critical rotation rate can now be defined as the value of $\Omega$ where the first and second term in Eq. (2) balance, i.e. $\Omega_c = T_0/K$. As shown in Fig 3b, this critical strain rate varies linear in the filling height, ruling out that the characteristic length scale ***L*** is the particle diameter.

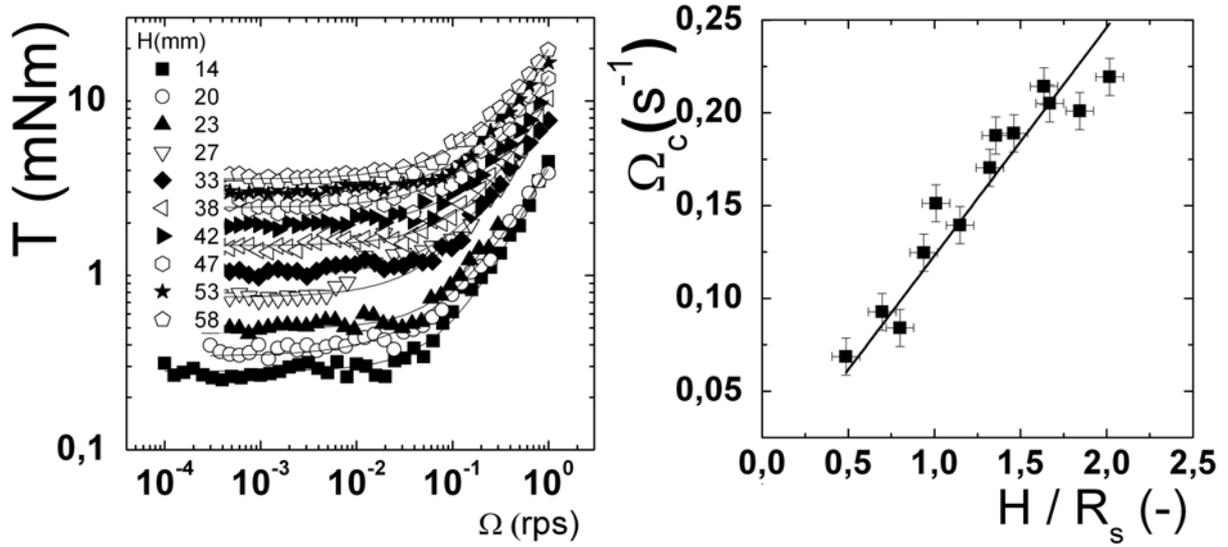

Fig. 3: Filling height dependence of the rheological measurements obtained in a Couette geometry with 4.6 mm diameter PMMA beads. a) Torque as a function of the rotation rate of the Couette cylinder. Different symbols (see the legend) correspond to different filling heights. Lines are fits to a Bingham model of the form $T = T_0 + K\Omega$: b) Variation of the critical rotation rate $\Omega_c$ as a function of filling height – here $\Omega_c = T_0/K$. The solid line is a linear fit of the form $\Omega_c = (0.12 \pm 0.02)\ H/R_s$.

### 3. A simple scaling model

The correct argument to obtain the critical shear rate is to focus on the balance of the static and dynamical stresses. Assuming an effective frictional picture to describe the rheological data, one can write: [Dijksman et al. 2010, GDRMidi 2004, Unger et al. 2004]

$$\sigma = \mu(I)P \approx [\mu_0 + \mu_1 I]P, \quad (4)$$

where $\mu_0$ and $\mu_1$ are the static and dynamic friction coefficients (presumably having a similar value), and $I$ is a dimensionless number capturing the effective friction in suspension rheology [Cassar et al. 2005]. The local static stress then equals $\mu_0 P$, and the dynamic stress $\mu_1 IP$ - using

$I = \frac{\eta \dot{\gamma}}{\alpha P}$, where $\alpha$ is a permeability parameter, we find that the dynamic stress is of order $\mu_1 \eta \dot{\gamma} / \alpha$ [Cassar et al. 2005].

To set up a proper balance of the total stresses, we need to integrate the static and dynamic stresses over the depth of the suspension bed, and use that the relevant pressure $P$ is the granular pressure due to gravity, $P = \Delta \rho g (H - z)$ suppose that the concentration profile is homogeneous so that $\Delta \rho$ is constant [Fall et al. 2009]. Equating the total static and dynamic stresses then yields:

$$\eta \dot{\gamma}_c = \Delta \rho g H \left( \frac{\alpha \mu_0}{2 \mu_1} \right) \qquad (3)$$

This can be compared, first, to the experimental data from the MRI rheology. The ratio $\mu_0 / \mu_1$ is of order unity, allowing us to extract the value $\alpha \approx 0.0013$ from a linear fit to the data shown in Fig. 2b. This estimate for $\alpha$ is in good agreement with the Carman-Kozeny equation [Carman, 1956], an extension of Darcy's law for flow through a porous medium that yields an explicit prediction for the proportionality constant between pressure drop and flow rate in Darcy's law:

$$\nabla p = \frac{45 U \eta}{R^2} \frac{(1-\varepsilon)^2}{\varepsilon^3}$$

where $\nabla p$ is the pressure gradient, $U$ the average fluid velocity and ε the porosity of the bed. In our case, the MRI results indicate that when the packing is dense, the porosity is very close to that for random close packing of spheres, so we will use ε = 0.36. Doing so, one obtains $\alpha \approx 0.00126$, in excellent agreement with our fitted value of α

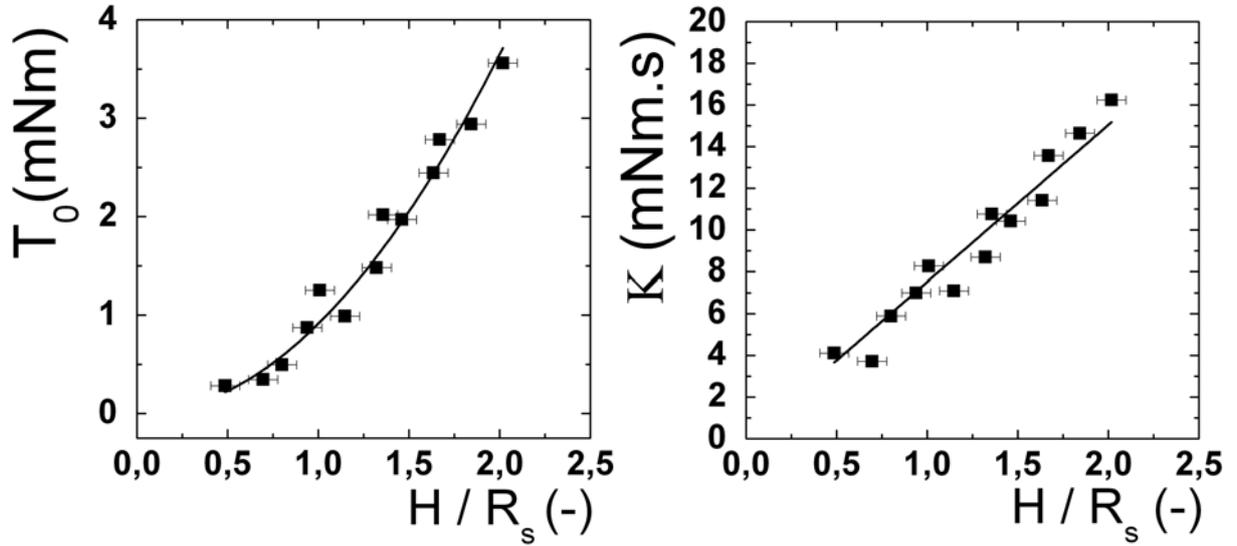

Fig. 4: a) Height dependence of the yield torque $T_0$. The line is a parabolic fit of the data of the form, as expected from Eq. 6. In good approximation, $T_0 = (0.9\pm0.1\ 10^{-3})$Nm $(H/R_s)^2$ b) Effective viscosity of the suspension as a function of the filling height. The solid line is a linear fit of the data of the form $K = (7.5 \pm 1\ .10^{-3})$ Nms $(H/R_s)$.

We can also compare the individual fitting constants **$T_0$** and **$K$** that arise in the Bingham fit to the rheological data of Fig 3a to theoretical predictions. For our Couette cell with inner radius **$R_s$**, outer radius **$R_0$**, and suspension filling height **$H$**, the torque is related to the stresses at the inner boundary as $T = 2\pi R_s^2 \int_0^H dz\, \sigma(z)$.

Substituting Eqs (4), using $\mu_1 IP = \frac{\mu_1 \eta \dot{\gamma} P}{\alpha P} = \frac{\mu_1 \eta \dot{\gamma}}{\alpha}$, using the equations for the hydrostatic pressure, and assuming that the shear rate is constant throughout the gap and given as $\dot{\gamma} = \Omega R_s / (R_0 - R_s)$, we obtain: $T = 2\pi R_s^2 [\int_0^H dz \{\mu_0 \Delta \rho g (H-z) + \frac{\mu_1 \eta}{\alpha}\dot{\gamma}\}]$, which performing the integrals yields:

$$T = \mu_0 \pi \Delta \rho g R_s^2 H^2 + 2\pi \frac{\mu_1 \eta}{\alpha} R_s^2 H \frac{R_s}{R_0 - R_s}\Omega \quad (5)$$

Identifying this expression with the Bingham fit, $T = T_0 + K\Omega$, we readily obtain:

$$T_0 = \mu_0 \Delta\rho g \pi R_s^4 \left(\frac{H}{R_s}\right)^2 \quad (6)$$

$$K = \mu_1 \frac{\eta}{\alpha} 2\pi R_s^3 \left(\frac{H}{R_s}\right)\left(\frac{R_s}{R_0 - R_s}\right) \quad (7)$$

This can be directly compared to the data. Fig. 4a indeed shows a quadratic increase of $T_0$ with (H/R$_s$), and substituting our experimental values in Eq.(6), we find that $T_0 \approx 10^{-3} \mu_0 (H/R_s)^2$ which is consistent with our fit shown in Fig. 4a, provided that $\mu_0$ is of order one. Similarly, we indeed observe a linear dependence of the 'effective viscosity' K with the filling height H in Fig. 4b, as predicted by Eq.(7), and substituting our experimental values in Eq.(7), we find that $K \approx 8.5\,10^{-3} \mu_1 (H/R_s)$, which is consistent with the fit shown in Fig. 4b, provided that $\mu_1$ is of order one. Finally, we can define a critical rotational rate $\Omega_c$ as the rotation rate for which 'frictional' and 'viscous' regime overlap, i.e. $T_0 = K\Omega_c$, and taking $\mu_0 = \mu_1$, it follows that :

$$\Omega_c = \frac{\mu_0}{\mu_1} \frac{\alpha \Delta\rho g H}{2\eta} \approx \dot{\gamma}_c \quad (8)$$

If we do so with a density difference of 110 kg/m$^3$, we obtain (from Fig. 2b) $\dot{\gamma}_c = 0.65\Omega_c$, in very reasonable agreement with the experimental observations. Thus, this simple scaling argument seems to be sufficient to account for all of the experimental observations presented in this paper.

4. **CONCLUSION.**

We have compared different experiments, using different techniques and performing the measurements on different systems to study the behavior of non-density matched non-Brownian suspensions. The absence of Brownian motion, together with the buoyancy and gravitational forces working on the particles make that large gradients in particle concentration develop in the direction parallel to gravity. The particles sediment or cream, which leads to a part of the system that is so dense in particles that it develops a yield stress. On the other hand, if the flows are sufficiently rapid, they homogenize the system. This naturally leads to the viscosity bifurcation, which in turn leads to the appearance of a critical shear rate above which the suspension is homogenized by the flow, and below which sedimentation or creaming happens. What exactly fixes the value of this critical shear rate was an outstanding question in the literature. Our experiments show that the value of the critical shear rate and its dependence on the experimental parameters are governed by a competition between the viscous and gravitational forces. This allows us to present a simple scaling model that predicts the value of the critical shear rate, as a function of the experimental parameters. It is found that this approach suffices to account for the experimental results from different types of experiments (local and global) in different setups and systems.

**Acknowledgements** we thank William Derek Updegraff from U. Maryland who performed the Triton Couette experiments. This work is part of the NWO-FOM programme "Jamming and Rheology".